\documentclass[aps,pra,twocolumn,showpacs,preprintnumbers,superscriptaddress,groupedaddress]{revtex4}
\usepackage{dsfont}
\usepackage{amsthm}
\usepackage{amsmath}
\usepackage{amssymb}
\usepackage{esint}
\usepackage{graphicx}
\usepackage{mathrsfs}             % See geometry.pdf to learn the layout options. There are lots.
\usepackage{graphicx}
\usepackage{amssymb}
\usepackage{epstopdf}
\usepackage{textpos}
\usepackage{color}
\usepackage{lipsum}
\usepackage{xcolor}
\begin{document}
\preprint{MIT-CTP/5135}
\title{Quantum Overlapping Tomography} % Parallelized / multiplexed? % Efficient?
\date{\today}

\author{Jordan Cotler}\email{jcotler@stanford.edu}

\affiliation{\it Stanford Institute for Theoretical Physics, Stanford University, Stanford, CA 94305, US}

\author{Frank Wilczek}\email{wilczek@mit.edu}

\affiliation{\it Center for Theoretical Physics, MIT, Cambridge MA 02139, USA}

\affiliation{\it T. D. Lee Institute, Shanghai, China}

\affiliation{\it Wilczek Quantum Center, Department of Physics and Astronomy, Shanghai Jiao Tong University, Shanghai 200240, China}

\affiliation{\it Department of Physics, Stockholm University, Stockholm Sweden}

\affiliation{\it Department of Physics and Origins Project, Arizona State University, Tempe AZ 25287 USA}

\begin{abstract}
It is now experimentally possible to entangle thousands of qubits, and efficiently measure each qubit in parallel in a distinct basis.  To fully characterize an unknown entangled state of $n$ qubits, one requires an exponential number of measurements in $n$, which is experimentally unfeasible even for modest system sizes.  By leveraging (i) that single-qubit measurements can be made in parallel, and (ii) the theory of perfect hash families, we show that \textit{all} $k$-qubit reduced density matrices of an $n$ qubit state can be determined with at most $e^{\mathcal{O}(k)} \log^2(n)$ rounds of parallel measurements.  We provide concrete measurement protocols which realize this bound.  As an example, we argue that with current experiments, the entanglement between every pair of qubits in a system of 1000 qubits could be measured and completely characterized in a few days.  This corresponds to completely characterizing entanglement of nearly half a million pairs of qubits.

\end{abstract}


\maketitle

%\emph{Introduction} ---

\section{Introduction}

Recently there have been remarkable advances in the construction and control of intermediate-scale quantum systems containing several hundred or even thousands of entangled qubits \cite{Ladd1, MonroeKim1, Devoret1, Awschalom1, 51qubit}.  The qubits come from a variety of systems, including interacting electronic spins, quantized fluxes, and spatial modes of photons.  But what about measuring the state of such systems, and documenting their entanglement?  

%Currently it is not possible to manipulate massively entangled states using universal quantum logic circuits with appreciable depth.    
To characterize an unknown $n$-qubit state completely using quantum tomography requires a number of parallel measurements which grows exponentially with $n$ \cite{O'Donnell1, Haah1}.  That exponential growth renders quantum tomography for many-body systems completely impractical even for modest system sizes.  Indeed, full quantum tomography has not been performed for more than 10 qubits \cite{Pan1}.  Some limited classes of quantum states featuring {\it a priori\/} constrained patterns of entanglement allow tomography with parametrically fewer measurements (for instance, see \cite{Cramer1, Lanyon1}), but most experimental systems do not produce states of those kinds.  There are ingenious protocols which can characterize expectation values of an unknown quantum state more efficiently \cite{Aaronson1}, but they require entangled non-demolition measurements and are not experimentally realistic for appreciably-sized systems.  Thus, there is a significant gap between our ability to produce massively entangled states in controlled settings, and our ability to  characterize that entanglement quantitatively.

What can be done is to address the individual qubits of a system in parallel, and to measure each in a chosen basis of $\mathbb{C}^2$.
Suppose we want to measure all $k$-qubit reduced density matrices of an $n$-qubit system.  Access to these density matrices would enable us to completely characterize all $k$-qubit entanglement present in the $n$-qubit system.  There are $\binom{n}{k}$ such $k$-qubit reduced density matrices, and if $k$ is small relative to $n$ then $\binom{n}{k} \sim n^k$.  Performing a $k$-qubit tomography requires $e^{\mathcal{O}(k)}$ measurements, and so na\"{i}vely we require $e^{\mathcal{O}(k)} \binom{n}{k} \sim e^{\mathcal{O}(k)} n^k$ measurements to obtain all $k$-qubit reduced density matrices.  Even for $k = 2$, it would not be practical to make so many measurements once $n$ exceeds a hundred qubits. 

This count, however, ignores the power of parallelism.   If we measure non-overlapping $k$-qubit subsystems in parallel we can get by with fewer measurements, but that only reduces the total number of required measurements by a multiplicative factor of $n/k$.  At first sight, it appears problematic that the set of all $k$-qubit subsystems is highly overlapping.  In fact, it is a tremendous advantage.  Measuring a particular $k$-qubit subsystem provides us information about all other $k$-qubit subsystems which overlap with it.    Here we present a method to organize that information.  We call it ``quantum overlapping tomography'' (QOT).   Using QOT, we can measure \textit{all} $k$-qubit reduced density matrices with at most $e^{\mathcal{O}(k)} \log^2(n)$ measurements. Our QOT protocols only require measuring each qubit in a distinct basis (i.e., a product measurement) in parallel, with judiciously chosen measurement settings.  The measurements can be efficiently post-processed to reconstruct all $k$-qubit reduced density matrices.  QOT easily adapts to qudits (i.e., $d$-level systems) in place of qubits.   $(n,k)$ families of perfect hash functions \cite{Mehlhorn1, Fredman1, Fredman2} will be a crucial tool in our measurement procedure.  The theory of perfect hash families has been well-studied in theoretical computer science for over thirty years, and is used in database management \cite{Mehlhorn1, Fredman1, Fredman2, Alon1, Korner1, Schmidt1, ColorCoding1, Naor1, Atici1, PerfectHashingReview1, Blackburn1, Blackburn2, Stinson1, Balanced1}.

We will begin by reviewing quantum tomography, and then provide a probabilistic argument for the scaling of our measurement procedure.  We then explain the measurement procedure in explicit mathematical detail for $k=2$, and more briefly for $k > 2$.  Then we describe its possible realization to measure all $2$-qubit entanglement in a system of ultracold atoms, and conclude with a summary and forward-looking discussion.

% \noindent $[[$Problem$]]$ \\
% $[[$quantum case harder; classical can read off whole configuration$]]$

\section{Review of Quantum Tomography}

Here we review the basic essentials of quantum tomography (see, for instance, \cite{Nielsen1}).  We focus on the standard experimental protocol which only requires product measurements, i.e.~measuring each qubit independently, since more sophisticated schemes involving entangled measurements are not presently experimentally feasible.  It will be useful to be very concrete about the measurement procedure.  To begin, we will explicitly explain how to do quantum tomography for a 2-qubit density matrix.

Suppose we have a 2-qubit density matrix $\rho$, and that we want to perform a quantum tomography of it.  To do so, we must be able to produce many copies of $\rho$, via some state preparation procedure, quantum source, etc.  Let $\sigma_i^\alpha$ denote a Pauli operator on the $i$th site, where here $i=1,2$ in the 2-qubit case.  We have $\alpha = 0,1,2,3$ where $\sigma^0 = \textbf{1}$, $\sigma^1 = \sigma^x$, $\sigma^2 = \sigma^y$, and $\sigma^3 = \sigma^z$, as is standard.  We can write $\rho$ as
\begin{equation}
\rho = \sum_{\alpha,\beta = 0}^3 \text{tr}\{(\sigma_1^\alpha \otimes \sigma_2^\beta) \, \rho \} \, \sigma_1^\alpha \otimes \sigma_2^\beta\,,
\end{equation}
and so to perform a quantum tomography we need to measure the expectation values $\text{tr}\{(\sigma_1^\alpha \otimes \sigma_2^\beta) \rho \}$, of which there are $4 \times 4 = 16$.  We do not need to measure the $\alpha = \beta = 0$ expectation value, since it is guaranteed to be $\text{tr}\{(\textbf{1} \otimes \textbf{1}) \rho \} = 1$ since $\rho$ has unit trace.  Thus, we only need to measure 15 expectation values.

First we consider the 1-site expectation values, for which either $\alpha = 0$, or $\beta = 0$.  For example, suppose we want to measure $\text{tr}\{(\sigma_1^0 \otimes \sigma_2^2)\rho\}$ where we recall that $\sigma_1^0 = \textbf{1}$.  Then we only need to measure the second qubit in the $y$-basis, and find that
\begin{equation}
\text{tr}\{(\sigma_1^0 \otimes \sigma_2^2)\rho\} \approx \frac{1}{M}\big(\mathcal{N}_2(\uparrow_y) - \mathcal{N}_2(\downarrow_y) \big)\,,
\end{equation}
where $\mathcal{N}_2(\uparrow_y)$ is the number of times we measure the second qubit to be up in the $y$-basis, and $\mathcal{N}_2(\downarrow_y)$ is defined similarly.
The other 1-site expectation values can be obtained in similar fashion.

Now we turn 2-site expectation values for which neither $\alpha$ nor $\beta$ equal zero.  As an example, considering the expectation value $\text{tr}\{(\sigma_1^1 \otimes \sigma_2^2)\rho\}$, we need to measure the first qubit in the $x$-basis, and concurrently the second qubit in the $y$-basis.  Let $\mathcal{N}_{12}(\uparrow_x\,, \uparrow_y)$ be the number of times we measure both the first qubit to be up in the $x$-basis, \textit{and} the second qubit to be up in the $y$-basis.  The quantities $\mathcal{N}_{12}(\uparrow_x\,, \downarrow_y)$, $\mathcal{N}_{12}(\downarrow_x\,, \uparrow_y)$, and $\mathcal{N}_{12}(\downarrow_x\,, \downarrow_y)$ are defined similarly.  If we make a total number of measurements $M$, then we can approximate
\begin{align}
\text{tr}\{(\sigma_1^1 \otimes \sigma_2^2)\rho\} \approx \, & \frac{1}{M}\big(\mathcal{N}_{12}(\uparrow_x\,, \uparrow_y) - \mathcal{N}_{12}(\uparrow_x\,, \downarrow_y) \nonumber \\
&\,\,\,\,\,\,\, - \mathcal{N}_{12}(\downarrow_x\,, \uparrow_y) + \mathcal{N}_{12}(\downarrow_x\,, \downarrow_y)\big)
\end{align}
which becomes exact in the limit of a large number of measurements $M$.  All other 2-site expectation values, for which neither $\alpha$ nor $\beta$ equal zero, can be obtained in an analogous manner.

Suppose we require $M$ measurements of each expectation value to obtain ample statistics.  If we want to do a tomography of $\rho$, which requires measuring 15 expectation values with $M$ measurements each, then na\"{i}vely we require $15 M$ measurements to determine $\rho$.  However, note that when we measure 2-site expectation values for which $\alpha$ and $\beta$ are both non-zero, we can use this data to extract 1-site expectation values.  For instance, upon collecting data to construct $\text{tr}\{(\sigma_1^1 \otimes \sigma_2^2)\rho\}$, we can use that same data to construct both $\text{tr}\{(\sigma_1^1 \otimes \textbf{1})\rho\}$ and $\text{tr}\{(\textbf{1} \otimes \sigma_2^2)\rho\}$.  Thus, instead of measuring all 15 expectation values to determine $\rho$, we effectively only need to measure 9 expectation values (i.e., the 2-site expectation values where neither $\alpha$ nor $\beta$ is zero), since we can reuse their measurements to reconstruct the other 6 expectation values.  In summary, we only require $9 M$ measurements to fully determine $\rho$. 

Now, suppose we have a $k$-qubit density matrix $\rho'$.  Writing $\rho'$ as
\begin{equation}
\rho' = \sum_{i_1,...,i_k = 0}^3 \text{tr}\{(\sigma_1^{i_1} \otimes \cdots \otimes \sigma_k^{i_k}) \, \rho' \} \, \sigma_1^{i_1} \otimes \cdots \otimes \sigma_k^{i_k}\,,
\end{equation}
we evidently need to determine $4^k - 1$ expectation values, where we have subtracted $1$ since we already know $\text{tr}\{(\textbf{1} \otimes \cdots \otimes \textbf{1}) \, \rho'\} = 1$.  Since we obtain each expectation value by multiplying the outputs of $k$ 2-outcome measurements, we need the probability that each measurement is faulty to be sufficiently small.  In particular, if the probability of a faulty measurement is $\Delta$, then we want $\Delta \sim 1/k$ so that $k \Delta \sim \mathcal{O}(1)$.

Using a similar procedure as in the 2-qubit case, we only need to perform $M 3^k$ total measurements, comprised of all combinations of $x$-basis, $y$-basis, and $z$-basis measurement settings for the $k$ sites, each repeated $M$ times to gain ample statistics.  If we want our approximations to all terms $\text{tr}\{(\sigma_1^{i_1} \otimes \cdots \otimes \sigma_k^{i_k}) \, \rho'\}$ to be within $\varepsilon$ of their true values with constant probability close to 1, then by the Chernoff-Hoeffding inequality and a union bound, we require $M$ to be at most $\sim k/\varepsilon^2$.  We will review the Chernoff-Hoeffding inequality in the Appendix.  In summary, we require $e^{\mathcal{O}(k)}$ measurements to perform a quantum tomography on $k$ qubits.

% As a technical aside, our reconstructed estimate of $\rho'$ may not exactly be a density matrix, i.e.~our reconstructed state may only be approximately positive semidefinite, and only approximately have unit trace.

\section{Probabilistic Argument}

In the last section, we saw that to perform a quantum tomography on $k$ qubits, we needed to perform measurements for all combinations of the measurement settings (either the $x$-basis, $y$-basis, or $z$-basis for each qubit), i.e.~varying the measurement basis of each qubit independently.  Since there are three bases for each qubit and $k$ total qubits, we required $3^k$ measurements, times a multiplicative factor of $M$ to build up enough statistics.

Given a system of $n$ qubits, we would like to measure all of its $k$-qubit reduced density matrices.  Defining $[n] := \{1,...,n\}$, we consider a surjective function $f : [n] \to [k]$ which assigns a number $1$ through $k$ to each qubit.  (Surjective means that for each number $1$ through $k$, there is at least one qubit assigned that value.)  Suppose we do a round of $M 3^k$ measurements as follows.  The function $f$ provides a partition of our system into $k$ sets $P_1,...,P_k$, where each set contains qubits assigned the same number by $f$.  For instance, $S_1$ contains all qubits assigned to the number $1$.  We pick a basis ($x$, $y$ or $z$) for each set, and measure \textit{all} qubits in that set in the selected basis.  For example, one parallel measurement may consist of measuring all $P_1$ qubits in the $x$-basis, all $P_2$ qubits in the $z$-basis, and so on.  There are $3^k$ ways of assigning measurement settings (i.e., a choice of basis) to the sets $P_1,...,P_k$, corresponding to $3^k$ parallel measurements.  We can repeat each set of parallel measurements $M$ times to gain statistics.

After these measurements, what have we learned?  Consider a $k$-qubit subsystem of the $n$ qubits, where each of the $k$ qubits was assigned to a distinct set $P_i$.  Then the aforementioned round of $M 3^k$ measurements is sufficient to determine the $k$-qubit reduced density matrix of such a subsystem.  More concretely, suppose for illustration that $|P_1| = |P_2| = \cdots = |P_k|$, meaning that $f$ equipartitions the $n$ qubits into $k$ sets of size $n/k$ each.  How many $k$-qubit subsystems have each qubit residing in a distinct set?  To construct such subsystems, we can choose one qubit from $P_1$, one qubit from $P_2$, and so on through $P_k$.  There are clearly $(n/k)^k$ combinations, and hence $(n/k)^k$ such subsystems.  Therefore, our $M 3^k$ measurements have allowed us to determine $(n/k)^k$ $k$-qubit density matrices!  To appreciate this, note that the na\"{i}ve parallelization strategy of concurrently performing quantum tomography on \textit{disjoint} $k$-qubit subsystems only allows us to learn $(n/k)$ $k$-qubit density matrices per $M 3^k$ measurements. 

Now we turn to constructing \textit{all} $\binom{n}{k}$ of the $k$-qubit reduced density matrices of the $n$-qubit system.  (Note that $(n/k)^k < \binom{n}{k}$, so we are not done yet.)  To formalize the problem, suppose we have a family of $N$ functions $f_1,...,f_N$, each taking $[n] \to [k]$.  These functions form an $(n,k)$ family of perfect hash functions if for any subset $S$ of $[n]$ where $|S| = k$ (i.e., $S$ contains $k$ elements), there there is some $f_i$ in the family which is injective on $S$  \cite{Mehlhorn1, Fredman1, Fredman2}.  For us, this means that for any given subsystem of $k$ qubits, there is at least one function $f_i$ in the family which assigns each qubit in that subsystem to a distinct number $1$ through $k$.

Given such a family of functions $f_1,...,f_N$, the approach of QOT is to run the procedure explained at the beginning of this section for each $f_i$.  This entails making a total of $N M 3^k$ total measurements, and allows us to determine all possible $k$-qubit reduced density matrices.  Then a crucial question is, what is the smallest $N$ for which we can construct an $(n,k)$ family of perfect hash functions?  

To construct a bound on $N$, we present a simple probabilistic argument, although there are more sophisticated bounds in the literature \cite{Fredman2, Korner1, Nilli1, Guruswami1}.  Suppose we choose each $f_i$ randomly, i.e.\! $f_i$ assigns each qubit to a number $1$ through $k$ uniformly at random.  We can ask: given $N$ random functions $f_1,...,f_N$, what is the probability that some subset $S$ of $[n]$ where $|S| = k$ has not been assigned in a 1-to-1 manner to $[k]$ by a function occurring so far?  

We proceed in steps.  Consider a particular subset $S$ of $[n]$ where $|S| = k$.  What is the probability that $f_1$ is 1-to-1 on $S$?  There are $k!$ ways to map each element of $S$ to a distinct element of $[k]$, and there are $k^k$ maps from $S \to [k]$.  So the probability that $f_1$ is 1-to-1 on $S$ is $k!/k^k$.  Therefore, the probability that $f_1$ is \textit{not} 1-to-1 on $S$ is $(1 - k!/k^k)$.  Then the probability that each of $f_1,...,f_N$ is not 1-to-1 on $S$ is $(1 - k!/k^k)^N$.  Finally, the probability that each of $f_1,...,f_N$ is not 1-to-1 on \textit{some} subset $S$ of $[n]$ of size $k$ is at most $\binom{n}{k} (1 - k!/k^k)^N$.  We would like this probability to be small, say less than some small parameter $\delta$\,:
\begin{equation}
\binom{n}{k} \left(1 - \frac{k!}{k^k} \right)^N < \delta\,.
\end{equation}
We immediately find that $N$ needs to be at most
\begin{equation}
\label{eq:bound1}
N < e^{\mathcal{O}(k)}\left( \frac{1}{k}\log(1/\delta) + \log(n)\right)
\end{equation}
which in turn implies that we require $M \, e^{\mathcal{O}(k)} \log(n)$ measurements to determine all $k$-qubit reduced density matrices of an $n$-qubit system using QOT.  Using the Chernoff-Hoeffding inequality and a union bound (see Appendix), if we want to determine all terms $\text{tr}\{(\sigma_1^{i_1} \otimes \cdots \otimes \sigma_k^{i_k}) \, \rho'\}$ within $\varepsilon$ of their true values with constant probability close to 1, then we require $M \sim k \log(n)/\varepsilon^2$.  Therefore, the total number of measurements is $e^{\mathcal{O}(k)} \log^2(n)$.

There is a substantial literature which constructs explicit and efficiently computable $(n,k)$ families of perfect hash functions which satisfy the bound in Eqn.~\eqref{eq:bound1}, such as \cite{Mehlhorn1, Fredman1, Fredman2, Alon1, Korner1, Schmidt1, ColorCoding1, Naor1, Atici1, PerfectHashingReview1, Blackburn1, Blackburn2, Stinson1, Balanced1}.  In the next section, we explain the simplest example, namely an explicit $(n,2)$ family of perfect hash functions of size $\lceil \log_2(n) \rceil$, which is well-known.

\section{QOT for $k = 2$}
\label{sec:keq2}

\begin{figure}[t]
  \centering
  \includegraphics[width=0.48\textwidth]{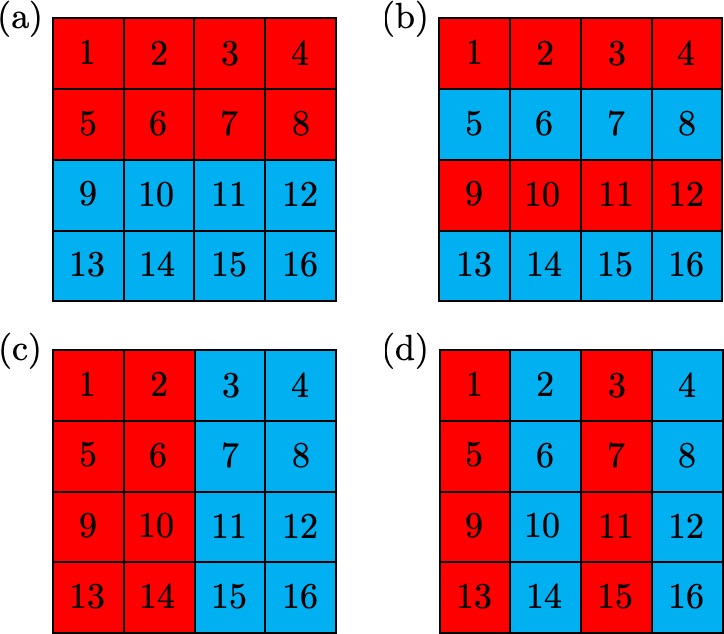}
    \vskip.2cm
  \caption{A visual depiction of the $(16,2)$ family of perfect hash functions given by $f_1, f_2, f_3, f_4$ from Eqn.~\eqref{eq:feq1}.  The four functions in the family are displayed in order in (a)--(d), where red corresponds to 0 and blue corresponds to 1.  Note that for any pair $(i,j)$ for $1 \leq i,j \leq 16$ and $i \not = j$, there is at least one function for which $i$ and $j$ are assigned distinct colors.}
  \label{fig1}
\end{figure}

In this section, we provide a QOT procedure for measuring all $2$-qubit reduced density matrices with only $(3M + 6 M \lceil \log_2(n) \rceil)$ measurements, for $M \sim 2\log(n)/\varepsilon^2$ as mentioned in the previous section.    We consider a simple but very useful example of an $(n,2)$ family of perfect hash functions, comprised of $q = \lceil \log_2(n) \rceil$ functions $f_1,...,f_q$ each taking $[n] \to \{0,1\}$.  (In our previous notation, we would have said that the functions take $[n] \to [2] = \{1,2\}$, but here we instead use $\{0,1\}$ as the codomain for convenience.)  The function $f_i$ is defined by
\begin{equation}
\label{eq:feq1}
f_i(j) = i\text{th digit in the binary expansion of }(j-1)\,.
\end{equation}
Here we are implicitly representing $(j-1)$ by a $q$-bit string, and by the $i$th digit we mean the $i$th most significant digit.  For instance, if we consider a $(16,2)$ family so that $q = 4$, then $f_1(5) = 0$, $f_2(5) = 1$, $f_3(5) = 0$ and $f_4(5) = 0$.  This follows from the fact that $4 = 5-1$ can be expressed as the $q$-bit string $0100$.  The functions $f_1, f_2, f_3, f_4$ are depicted in Fig.~\ref{fig1}.

Suppose that we have $n$ qubits, and that we want to perform quantum tomography on every 2-qubit reduced density matrix using QOT.  We consider an $(n,2)$ family of perfect hash functions given by Eqn.~\eqref{eq:feq1} with $q = \lceil \log_2(n) \rceil$.  The procedure is as follows:
\\ \\
\textbf{Step 1:} Measure all qubits in the $x$-basis, $y$-basis, and $z$-basis, each $M$ times.  Since all of the qubits can be measured in parallel, this corresponds to $3M$ measurements.
\\ \\
\textbf{Step 2:} This step will be divided into $q$ substeps, $2.1,...,2.q$.  For each $j = 1,...,q$, Step $2.j$ is as follows.  Consider the function $f_j$.  If a qubit is assigned to $0$ by $f_j$, then we call the qubit ``red''.  Similarly, if a qubit is assigned to $1$ by $f_j$, then we call the qubit ``blue''.  Then we perform the following 9 measurements, $M$ times each:
\begin{itemize}
\item Measure each red qubit in the $\mathbf{x}$\textbf{-basis}, and each blue qubit in the $\mathbf{y}$\textbf{-basis}. 
\item Measure each red qubit in the $\mathbf{y}$\textbf{-basis}, and each blue qubit in the $\mathbf{x}$\textbf{-basis}.
\item Measure each red qubit in the $\mathbf{x}$\textbf{-basis}, and each blue qubit in the $\mathbf{z}$\textbf{-basis}.
\item Measure each red qubit in the $\mathbf{z}$\textbf{-basis}, and each blue qubit in the $\mathbf{x}$\textbf{-basis}.  
\item Measure each red qubit in the $\mathbf{y}$\textbf{-basis}, and each blue qubit in the $\mathbf{z}$\textbf{-basis}. 
\item Measure each red qubit in the $\mathbf{z}$\textbf{-basis}, and each blue qubit in the $\mathbf{y}$\textbf{-basis}. 
\end{itemize}
Due to parallelization, each Step 2.$j$ corresponds to $6 M$ measurements, and thus $6 M q = 6 M \lceil \log_2(n) \rceil$ measurements total for all of Step 2. \\

\noindent \textbf{Step 3:} Steps 1 and 2 collect all of the data we need, and only require a total of $3M + 6 M \lceil \log_2(n) \rceil$ measurements.  Suppose we want to reconstruct the reduced density matrix $\rho_{rs}$ of the $r$th qubit and the $s$th qubit, for $1 \leq r,s \leq n$ and of course $r \not = s$.  Note that the $q$-bit binary representation of $(r-1)$ and $(s-1)$ must differ on at least one bit, since $r$ and $s$ are distinct numbers.  Suppose that $(r-1)$ and $(s-1)$ differ on their $t$th bits.  Then:
\begin{itemize}
\item To obtain $\text{tr}\{(\textbf{1}_r \otimes \sigma_s^x) \rho_{rs} \}$, $\text{tr}\{(\textbf{1}_r  \otimes \sigma_s^y) \rho_{rs} \}$, $\text{tr}\{(\textbf{1}_r  \otimes \sigma_s^z) \rho_{rs} \}$, $\text{tr}\{(\sigma_r^x \otimes \textbf{1}_s ) \rho_{rs} \}$, $\text{tr}\{(\sigma_r^y \otimes \textbf{1}_s) \rho_{rs} \}$, $\text{tr}\{(\sigma_r^z \otimes \textbf{1}_s) \rho_{rs} \}$, we use the data collected from Steps 1 and 2.
\item To obtain $\text{tr}\{(\sigma_r^x \otimes \sigma_s^x) \rho_{rs} \}$, $\text{tr}\{(\sigma_r^y \otimes \sigma_s^y) \rho_{rs} \}$, $\text{tr}\{(\sigma_r^z \otimes \sigma_s^z) \rho_{rs} \}$, we use the data collected from Step 1.
\item To obtain $\text{tr}\{(\sigma_r^x \otimes \sigma_s^y) \rho_{rs} \}$, $\text{tr}\{(\sigma_r^y \otimes \sigma_s^x) \rho_{rs} \}$, $\text{tr}\{(\sigma_r^x \otimes \sigma_s^z) \rho_{rs} \}$, $\text{tr}\{(\sigma_r^z \otimes \sigma_s^x) \rho_{rs} \}$, $\text{tr}\{(\sigma_r^y \otimes \sigma_s^z) \rho_{rs} \}$, $\text{tr}\{(\sigma_r^z \otimes \sigma_s^y) \rho_{rs} \}$, we use the data collected from Step $2.t$.
\end{itemize}
Then we can reconstruct $\rho_{rs}$ using
\begin{equation}
\rho_{rs}= \sum_{\alpha,\beta = 0}^3 \text{tr}\{(\sigma_r^\alpha \otimes \sigma_s^\beta) \, \rho_{rs} \} \, \sigma_r^\alpha \otimes \sigma_s^\beta\,.
\end{equation}
Once we have all of the 2-qubit reduced density matrices at hand, we can analyze their bipartite entanglement.  For instance, there are explicit formulas for computing the entanglement of formation \cite{Wooters1} and related quantities \cite{Vidal1, Page1}.  One can then study, for example, how entanglement varies as the qubits comprising the 2-qubit subsystem are chosen to be further apart in space.

\section{QOT for arbitrary $k$}

To perform QOT to determine all $k$-qubit reduced density matrices of an $n$-qubit system, one proceeds in the same way as in the previous section, but instead utilizing an $(n,k)$ family of perfect hash functions.  In the language of the previous section, each function $f_i$ in the family assigns each qubit to one of $k$ ``colors'', i.e.\! red, blue, green, etc.  The procedure generalizes in the obvious way.  Then the total number of required measurements scales as $M \, e^{\mathcal{O}(k)} \log(n) \sim e^{\mathcal{O}(k)} \log^2(n)$, which has an $n$-dependence significantly better than even shadow tomography applied to measuring subsystems \cite{Aaronson1}.  Such a shadow tomography would require $\mathcal{O}(n \, \text{polylog}(n))$ measurements.

For $k>2$, constructing $(n,k)$ families of perfect hash functions which contain as few functions as possible can be a difficult task.  Luckily, there is an extensive literature on constructing such families, and we refer the reader to \cite{Mehlhorn1, Fredman1, Fredman2, Alon1, Korner1, Schmidt1, ColorCoding1, Naor1, Atici1, PerfectHashingReview1, Blackburn1, Blackburn2, Stinson1, Balanced1}.  Also, there is a web page providing a list of the smallest known $(n,k)$ families for various values of $n$ and $k$ \cite{FamilyList1}.

Let us make several comments.  First, consider an algorithm, such as \cite{ColorCoding1}, for efficiently constructing $(n,k)$ families of perfect hash functions with at most $e^{\mathcal{O}(k)} \log(n)$ functions.  (It is known how to do slightly better than this asymptotically; for instance, see \cite{Naor1}.)  Note that the bound $e^{\mathcal{O}(k)} \log(n)$ on the number of functions is asymptotic, and so there is no guarantee on the optimality of the size of the hash family for fixed finite values of $n$ and $k$.
%\textcolor{red}{[CHECK THIS]} 
For QOT, we ideally desire $(n,k)$ families of perfect hash functions which contain as few functions as possible, for \textit{particular} values of $n$ and $k$.  Of course, suboptimal constructions of families suffice in practice, although they entail making more measurements than is in principle necessary. 

While it is sensible to find $(n,k)$ families which are as small as possible, there are circumstances in which other properties are desirable.  As an example, note that for an $(n,k)$ perfect family $f_1,...,f_N$, for a given $S \subset [n]$ with $|S| = k$, we are only guaranteed that at least one $f_i$ is 1-to-1 on $S$.  However, suppose that $T$ of the functions are 1-to-1 on $S$.  If this was the case for \textit{every} $|S| = k$, then we could reduce the number of measurement repetitions from $M$ to $M/T$.  So instead of requiring $M \, e^{\mathcal{O}(k)} \log(n)$ measurements (here we are being more explicit about the $M$-dependence), we would only require $(M/T) \, e^{\mathcal{O}(k)} \log(n)$ measurements.  Finding $(n,k)$ families with this $T$-property is difficult, but there is a useful approximate notion.  A $(\delta, T)$--balanced $(n,k)$ family of perfect hash functions $f_1,...,f_N$ has the property that for every $S$ in $[n]$ where $|S| = k$, there are between $T/\delta$ and $\delta T$ functions which are 1-to-1 on $S$  \cite{Balanced1}.  For any given $\delta > 1$,  \cite{Balanced1} provides a construction of a $(\delta, T)$--balanced $(n,k)$ family of size $e^{\mathcal{O}(k \log \log (k))} \log(n)$, where $T$ is determined by the construction.  This would allow for QOT with $(M/T) \, e^{\mathcal{O}(k \log \log (k))} \log(n)$ measurements, which in certain parameter regimes would require less measurements than the non-balanced case.

A different generalization involves $(n,t,k)$ families of perfect hash functions, where $n \geq t \geq k$.  These are a family of functions $f_1,...,f_N$, each taking $[n] \to [t]$, such that for any subset $S$ of $[n]$ where $|S| = t$, there is some $f_i$ which is injective on $S$.  (That is, $f_i$ maps each element of $S$ to a different element of $[t]$.)  The $(n,k)$ families previously described correspond to $(n,k,k)$ families of this more general kind.  In QOT, the number of required measurements scales exponentially with $t$, and so it appears that we should choose $t$ as small as possible, namely to be $k$.  The brings us back to the $(n,k)$ perfect hash families discussed above.  However, for (i) particular values of $(n,t,k)$, or (ii) if we are performing QOT with a restricted class of all $k$-site expectation values which requires less than $e^{\mathcal{O}(t)}$ measurements, then it may be advantageous to leverage $(n,t,k)$ families to reduce the total number of measurements.

% \textcolor{red}{[go over notation and comments carefully here -- especially for $(n,t,k)$ families]}

%\noindent $[[$ schematically $]]$ \\
%$[[$ $\mathcal{O}(1)$ factors make a difference for an experimentalist $]]$

%\section{Geometric variation}
%\noindent $[[$ probabilistic argument $]]$ \\
%$[[ \, k = 2 \, ]]$ \\
%$[[$ higher $k$ $]]$ \\
%$[[$ nearest-neighbor on a rectangular lattice requires $\mathcal{O}(1)$ measurements -- done before \textcolor{red}{[CITE]} $]]$

\section{Experimental Prospects}

Here we estimate the practical potential of QOT based on currently attainable technology.  Consider an ultracold atom system with spin-1/2 degrees of freedom, which we can prepare in the ground state of a local Hamiltonian and then probe with a quantum gas microscope (for a review, see \cite{Kuhr1}).  In arrays of neutral atoms configured using optical tweezers \cite{51qubit, Barredo1} or in arrays of optically trapped ions \cite{53qubit}, each measurement round takes at most a few hundred milliseconds.  A single cycle of the experimental protocol can be significantly faster, even for systems sizes up to hundreds or even thousands of atoms.  
Concretely, let us choose $n = 1024$ qubits, subsystems of size $k=2$, and $M = 16,\!000$, so that around 97\% of the time \textit{all} measured expectation values are within 0.05 of their true values.  (See Appendix for more details on the estimation of $M$.) Then the $k=2$ QOT protocol in Section~\ref{sec:keq2} requires $1,\!000,\!800$ measurement rounds.  Assuming 250 milliseconds per measurement round, $k=2$ QOT could be performed in a block of 3 days.  By contrast, the na\"{i}ve strategy of measuring non-overlapping $2$-qubit subsystems in parallel with similar error probabilities requires $M = 5,\!500$ and thus $9M \binom{n}{2}/(n/2) = 50,\!638,\!500$ total measurement rounds, which would take nearly 21 weeks nonstop.  Thus, QOT would enable measurement of entanglement between $499,\!500$ pairs of qubits comprising a many-body quantum state.  (It would be especially interesting to use 2D or 3D arrays, since in the higher-dimensional setting it is easier to trap a large number of atoms, and also more interesting to characterize low-energy eigenstates of experimentally realizable Hamiltonians.)   Even for $n$ of 1000 or more, it is possible using QOT to measure \textit{every} 2-qubit reduced density matrix $\rho_{rs}$, and thus to characterize their entanglement precisely \cite{Wooters1}.  One expects that in the ground state many pairs will be entangled.

% Get inspiration from Review (ref 1, ref 3, ref 4, ... )
% Quantum thermalization paper Greiner
% Bakr paper (?)
% Immanuel Bloch
% Pan... (?)
% Comment on 2D arrays vs. 1D arrays
% Rydberg atoms or trapped ions -- 250 ms! -- 51 qubit simulator
% Trapped ion or Rydberg atom array, each measurements takes a few hundred milliseconds [51 qubit simulator, 53 qubit simulator] less than 1 second
% Reconfigurable arrays of neutral atoms in optical tweezers demonstrated in [51 + french]
% A single cycle of the experimental protocol can be significantly faster, even for systems sizes up to hundreds or even thousands of atoms.

\section{Summary and Discussion}

QOT provides efficient protocols to measure many-body correlations and entanglement in systems with large numbers of degrees of freedom.  We anticipate that QOT will be a useful tool for experimental characterization of many-body quantum states.  

Several adaptations of QOT may be interesting to consider.  Systems with symmetry obey constraints and selection rules which might be exploited to streamline the protocol. One might also try to focus on local entanglement, in systems where long-range entanglement is not significant.  This poses interesting mathematical problems.  
For example, given an $n$-qubit state on a lattice, how do we efficiently measure all $k$-qubit reduced density matrices, for $k$-qubit subsystems where every pair of qubits is at most a distance $d$ apart?  Taking geometric constraints into account would require a generalization of $(n,k)$ families of perfect hash functions, entailing the additional data of (i) a weighted graph $G$ representing the geometry, and (ii) a distance $d$ which serves as the maximum diameter of the $k$-qubit subsystems.  As one example, if we wanted to measure all nearest-neighbor correlation functions (i.e., $d=2$) of an $n$-qubit system on a square lattice, then we only require $M$ measurements.  The corresponding ``coloring'' of the $n$ qubits is to color every other site as red, and color the remaining sites blue, forming an alternating checkerboard pattern.

% This special case has been implemented in experiments already. 

Since QOT allows us to efficiently measure \textit{all} $k$-point functions of a system, it would be natural to use QOT to diagnose long-range order and critical behavior.  A modification of the QOT protocols may be useful to focus on special types of non-local  order parameters (for instance, string-like products) which appear in the classification of topological order (see e.g. \cite{Wen1, Wen2, Wen3}).

QOT can be applied to measuring expectation values of $k$-local Hamiltonians, such as those which appear in quantum and classical versions of $k$-SAT \cite{Kitaev1, Kitaev2} and in recent work on quantum machine learning \cite{QML1, QML2, QML3, QML4, QML5, QML6, QML7}.  Also, QOT can supply needed input for the quantum marginal problem (see \cite{Walter1, Schilling1} for recent overviews, and \cite{Xin1} for applications in tomography), in which one tries to determine a quantum state as well as possible given its reduced density matrices up to a given size.

Finally, one should be able to adapt QOT to quantum channel tomography.  There has been recent work in the direction of diagnosing quantum channels via $k$-point marginals \cite{Flammia1, Flammia2}, and so QOT may be useful in this context. \\

\noindent \textbf{Acknowledgments.\quad} A special thank you to Jian-Wei Pan and Yu-Ao Chen for suggesting the problem and discussing experimental implementations.  JC is grateful to Reuben Saunders for earlier discussions on multiplexed RNA perturb-sequencing.  We are happy to thank Noga Alon, Ryan Alweiss, and Xiaoyu He for suggesting valuable references on perfect hash functions, Soonwon Choi for guidance on ultracold atom references, and Patrick Hayden, Steve Flammia, and Daniel Ranard for discussions and feedback on the manuscript.  JC is supported by the Fannie and John Hertz Foundation and the Stanford Graduate Fellowship program.  FW's work is supported by the U.S. Department of Energy under grant Contract  Number DE-SC0012567, by the European 
Research Council under grant 742104, and by the Swedish Research Council under Contract No. 335-2014-7424.

\cleardoublepage
% \begin{widetext}
\onecolumngrid
\section*{Appendix: Application of the Chernoff-Hoeffding Inequality}
\appendix
One version of the Chernoff-Hoeffding inequality is as follows.  Given $M$ i.i.d. random variables $X_j$, each valued on $[a,b]$, let $Y = \frac{1}{M}\sum_{j=1}^M X_j$.  Then
\begin{equation}
\label{eq:CHinequality}
\text{Pr}\left[ \, |Y - \mathbb{E}[Y]| > \varepsilon \right] \leq 2 \exp\left(- \frac{2 M \, \varepsilon^2}{(b-a)^2}\right)\,.
\end{equation}
In words, the probability that $Y$ deviates from its expected value $\mathbb{E}[Y]$ by more than $\varepsilon$ is exponentially suppressed in $M \, \varepsilon^2$.

In the setting of this paper, we want to measure expectation values $\text{tr}\{(\sigma_1^{i_1} \otimes \cdots \otimes \sigma_k^{i_k}) \, \rho' \}$ for each of $\binom{n}{k}$ $k$-qubit subsystems.  The expectation values are each valued in $[-1,1]$.  There are $(4^k - 1)\binom{n}{k}$ such (non-trivial) expectation values.  Let $X_j^{(i)}$ for $i=1,...,(4^k - 1)\binom{n}{k}$ and $j=1,...,M$ denote the outcome of a measurement of one of the expectation values (i.e., the $i$th one) during the $j$th round of measurement.  So our estimate of the $i$th expectation value is $Y^{(i)} = \frac{1}{M}\sum_{j=1}^M X_j^{(i)}$.  Note that by our measurement protocol, each $X_j^{(i)}$ is valued in the discrete set $\{-1,1\}$, and thus also lives in the interval $[-1,1]$.  Also, for fixed $i$, the random variables $X_1^{(i)},...,X_M^{(i)}$ are i.i.d. since they correspond to outcomes of independent, sequential measurements.  On the other hand, for fixed $j$, the random variables $X_{j}^{(i)}$ and $X_{j}^{(i')}$ for $i \not = i'$ will \textit{not} be i.i.d. if they correspond to overlapping correlation functions. 

Using the Chernoff-Hoeffding inequality in Eqn.~\eqref{eq:CHinequality}, we find that our estimate of the $i$th expectation value (for $i=1,...,(4^k - 1)\binom{n}{k}$\,) is
\begin{equation}
\label{eq:CHinequality2}
\text{Pr}\left[ \, |Y^{(i)} - \mathbb{E}[Y^{(i)}]| > \varepsilon \right] \leq 2 \exp\left(- \frac{M \, \varepsilon^2}{2}\right)\,.
\end{equation}
and by a union bound
\begin{align}
\text{Pr}\left[ \, |Y^{(i)} - \mathbb{E}[Y^{(i)}]| > \varepsilon\,\,\, \text{for all }i \,\right] &\leq \sum_{i=1}^{(4^k - 1)\binom{n}{k}}\text{Pr}\left[ \, |Y^{(i)} - \mathbb{E}[Y^{(i)}]| > \varepsilon \right] \\
&\leq 2 (4^k - 1) \binom{n}{k}  \exp\left(- \frac{M \, \varepsilon^2}{2}\right)\,.
\end{align}
% &\leq (4n)^k  \exp\left(- \frac{M \, \varepsilon^2}{2}\right)\,.
If we want the right-hand side to be at most some small number $\delta$, then we set
\begin{equation}
M = \frac{2}{\varepsilon^2} \, \log\left(2\,(4^k - 1)\binom{n}{k} \right) \sim \frac{1}{\varepsilon^2} \, k \log(n)\,.
\end{equation}

%\cleardoublepage
%\onecolumngrid
%\section*{Supplementary Materials}
%\appendix
%\section*{I. Further details...}

\end{document}